\documentclass[11pt]{article}
\newsavebox{\PSLASH}
\sbox{\PSLASH}{$p$\hspace{-1.8mm}/}
\def\d{\partial}
\def\db{\bar\partial}
\def\t{\theta}
\def\tb{\bar\theta}

\begin{document}
\title{Abelian Sandpile Model: a Conformal Field Theory Point of View}
\author{S. Moghimi-Araghi \footnote{e-mail:
samanimi@sharif.edu} , M. A. Rajabpour \footnote{e-mail:
rajabpour@mehr.sharif.edu}
, S. Rouhani \footnote{e-mail: rouhani@ipm.ir} \\ \\
Department of Physics, Sharif University of Technology,\\ Tehran,
P.O.Box: 11365-9161, Iran} \maketitle

\begin{abstract}
In this paper we derive the scaling fields in $c=-2$ conformal
field theory associated with weakly allowed clusters in abelian
sandpile model and show a direct relation between the two models.
\end{abstract}

\section{Introduction}
There exists some phenomena which naturally show power law
behavior, that is, without fine tuning any parameter, the system
shows behavior similar to the critical point, in contrast with the
usual critical phenomena, where you should fine tune an external
parameter like temperature to arrive at the critical point. These
kind of phenomena are said to have self organized criticality
\cite{BTW}. Sandpile \cite{BTW,Dhar1}, surface growth
\cite{StanBarb} and river networks \cite{River} are a few examples
of such phenomena.

The concept of SOC was first introduced by Bak, Tang and
Wiesenfeld \cite{BTW}. While many other models have been found
after that, still the abelian sandpile model (ASM) is one of the
simplest and most studied models. Despite its simplicity, it shows
all the features that a self organized critical phenomena ought to
present, so huge amount of work has been done on this model
[5-18]. Many analytic results have been derived, for example the
probability of different heights and many specific clusters are
calculated explicitly \cite{MajumDhar,Priez}. Also the relation of
this model to other known models has been discussed. First of all,
there is a connection between ASM and spanning trees \cite{MD}.
Then other models such as dense polymers, Scheidegger's model of
river networks and $q\rightarrow 0$ limit of $q$-state Potts
models arise\cite{Saluer}.

These statistical models, show conformal symmetry while they are
at their critical point. So it is natural to look for a conformal
field theory (CFT) which corresponds to these systems. The CFT
associated with these models is suggested to be $c=-2$, which
belongs to a specific group of CFT's, known as logarithmic
conformal field theories (LCFT's). In LCFT's, where correlation
functions may have logarithmic terms in contrast with the ordinary
CFT's, there exist pairs of fields with the same conformal weight,
which mix under conformal transformations\cite{Gur,Flohr}.

Maheiu and Ruelle \cite{MaRu} have found a way to relate ASM to
$c=-2$ theory. They have found some operators in the LCFT model
which correspond to different clusters in ASM. But the
correspondence is shown only through correlation function and it
is not clear why one should take these operator. In this paper
we'll address this question and find a direct way to derive the
operators from the action of $c=-2$, and hence connect ASM to
$c=-2$ directly. The paper is organized as follows: in section 2
we will briefly introduce the model and some analytical results
including the result obtained by \cite{MaRu}. In the 3rd section
we will derive the previously mentioned operators directly from
the action of $c=-2$ model.

\section{Abelian Sandpile Model}

The Abelian Sandpile Model is defined on a square lattice of the
size $L\times M$. To each site $i$, a height variable $h_i$ is
assigned. Height of sand at each site can take one of the values
form the set $\{ 1,2,3,4 \}$. So the total number of different
allowed configuration is equal to $4^{L\,M}$. The dynamic of the
system is defined as follows: at each step a random site $i$ is
selected and a grain of sand is added to that site. If the new
height of sand becomes more than four, the column of sand is
called to be unstable and topples, that is, four grains will leave
the site and each of them will be added to one of the neighbors.
So the total number of sands is conserved during the toppling
process except at the boundaries where one or more grains leave
the system.

The toppling process can be stated in another way, which will be
more appropriate. If the site $i$ becomes unstable, $h_j$ will be
decreased by amount of $\Delta_{ij}$, that is $h_{j}\rightarrow
h_{j} - \Delta_{ij}$ where
\begin{eqnarray}\
    \Delta_{ij}=\left\{%
\begin{array}{ll}
    4, & \hbox{ $i=j$;} \\
    -1, & \hbox{ $i, j$ are neighbors;} \\
    0, & \hbox{ otherwise.} \\
\end{array}%
\right.
\end{eqnarray}
The matrix $\Delta_{ij}$ is called toppling matrix.

After a while the system reaches a steady state, in which it shows
SOC. It has been shown that in this state, the number of different
configurations the system accepts is $\det \Delta$ and the
probability of all of them are the same. These configurations are
named recurrent configurations in contrast with the transient ones
which can only appear in the first steps of evolution, where the
system has not yet reached the steady state. Note that the number
of recurrent configurations is fairly smaller than the total
number of configuration as $\det \Delta \sim 3.2^{L\,M}$.
Determining whether a given configuration is recurrent or
transient is a relatively hard question, though there exist tests
to answer this question. The first observation is that some
forbidden subconfigurations exist, that is if in a given
configuration you find one of these subconfigurations then it
could not be recurrent. The simplest example of these
subconfigurations is two neighboring height one sites.

One of the most important analytic results derived so far, is the
probability of finding some specific clusters in the system while
it is in the steady state. Dhar and Majumdar \cite{MajumDhar} have
found a subtle way to calculate such probabilities for the
clusters named weakly allowed clusters (WAC's). WAC's are those
clusters which by decreasing the height of any of its sites by
one, it becomes a forbidden subconfiguration. The simplest one is
a single site with height equal to one. They have proved that the
number of recurrent configuration with a particular WAC is equal
to the number of recurrent configurations in another sandpile
model with a modified toppling rule. In fact there are several
different ways to modify the toppling rule. One of them is to
remove all the connections of the WAC to the rest of the system
but one as shown in figure 1 for two simplest clusters known as
$S_0$ and $S_1$ respectively\footnote{If you are considering
several disconnected pieces of WAC's, you should do the same for
each piece.}. As the grains of sand are not allowed to flow
through the disconnected bonds, both the condition for instability
and the toppling matrix should be modified. The new toppling
matrix is given by $\Delta'=\Delta+B$, where for each disconnected
bond running from site $i$ to site $j$, $B_{ij}=B_{ji}=1$ and if
$n$ bonds have been cut from site $i$ we should set $B_{ii}=-n$.
As it is clear, the matrix $B$ is nonzero only in the vicinity of
the cluster we are dealing with. So, $B$ is somehow local.
\begin{figure}[t]
\begin{picture}(200,200)(0,0)
\includegraphics{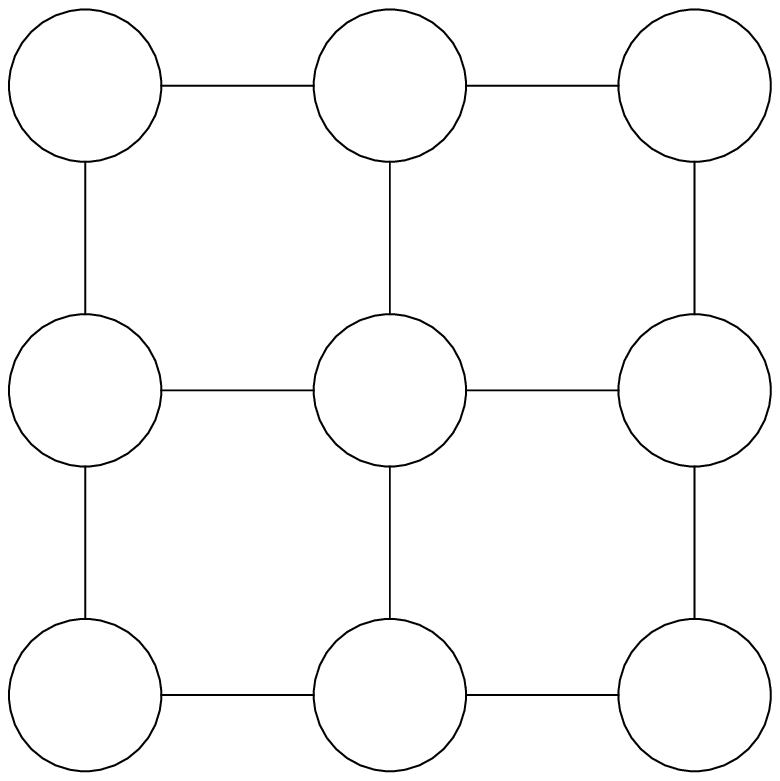} \includegraphics{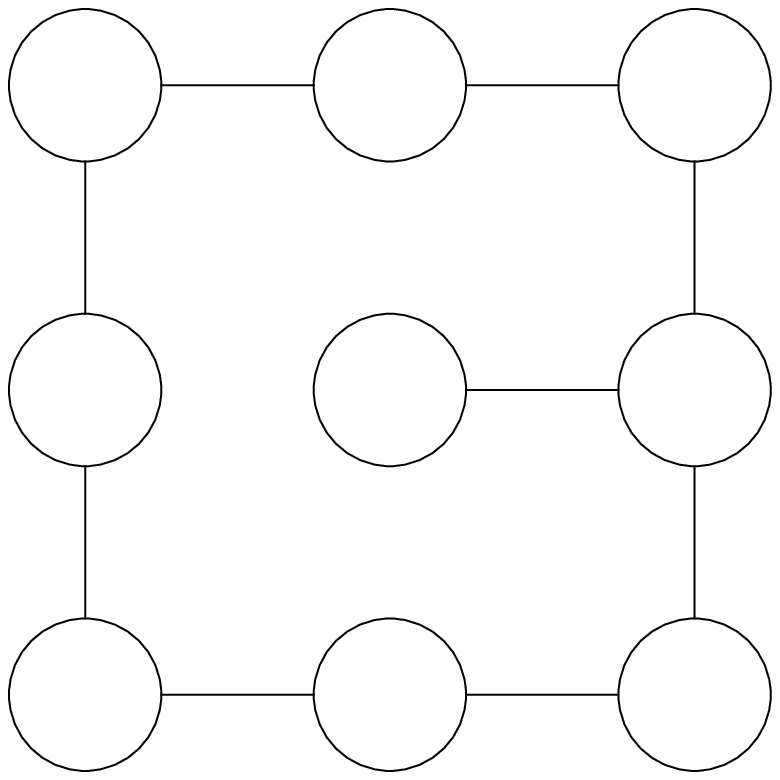}\includegraphics{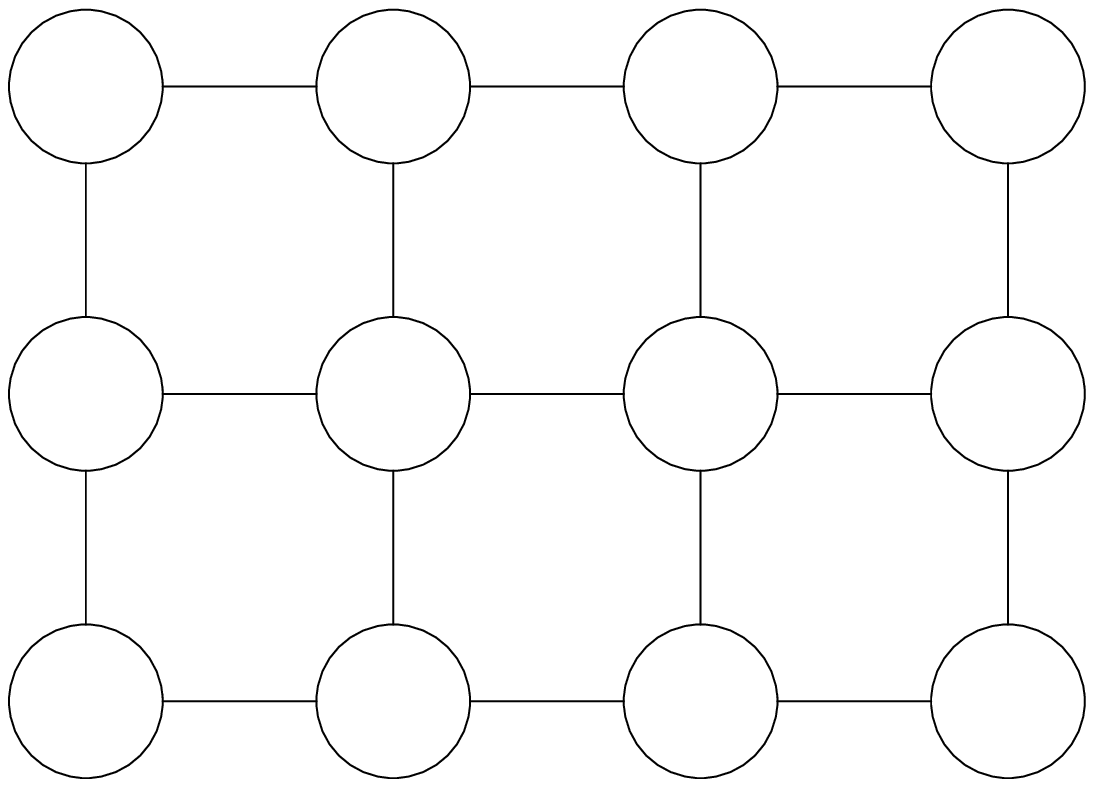} \includegraphics{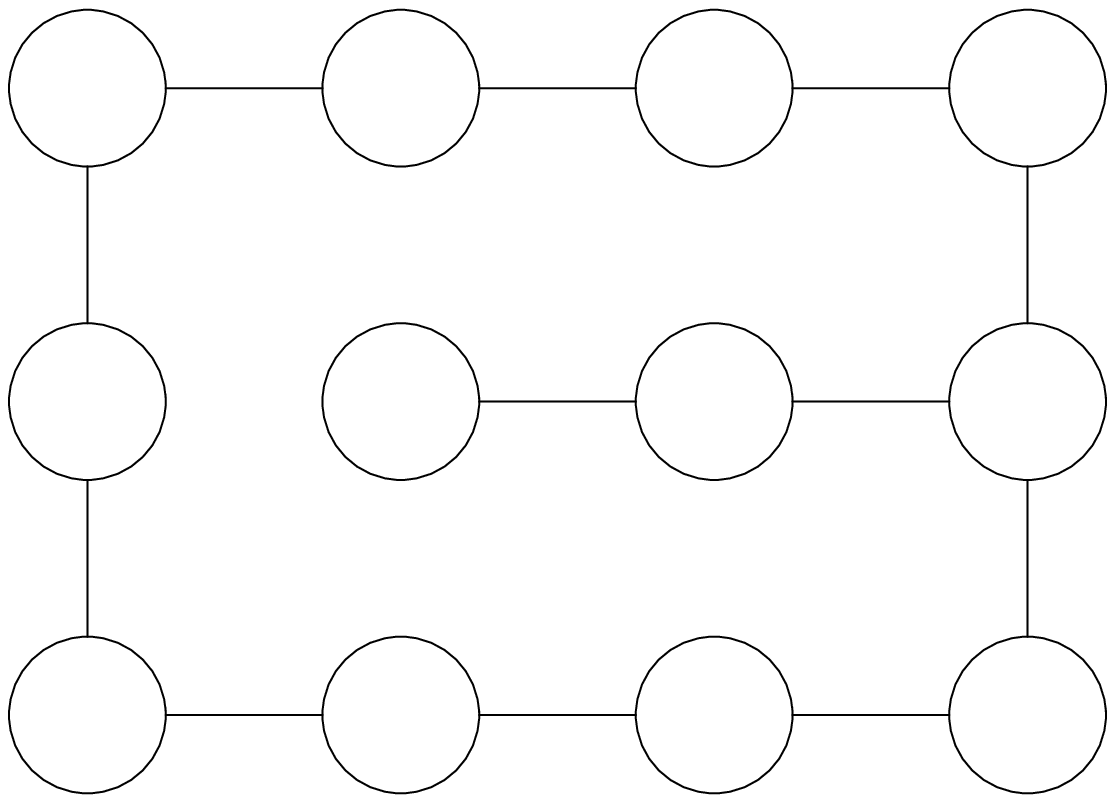}
\includegraphics{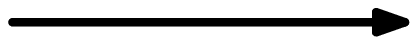}\includegraphics{arrow.eps} \put(99.5,185){$1$}\put(304,184.5){$1$}
\put(94.5,65){$1$}\put(299.5,65.5){$1$}\put(130,65){$2$}\put(335,65.5){$2$}
\end{picture}
\caption{Modification of the sandpile lattice in the two simplest
Clusters $S_0$ and $S_1$}
\end{figure}

There are some other ways to modify the lattice and the toppling
matrix. One of them which is more appropriate for our future uses,
is to disconnect the cluster completely from other sites of
lattices. This scheme is better because it preserves the
symmetries of the clusters manifestly. The rules for modification
of the toppling matrix is more or less the same as before, the
only thing you should take into account is that $B_{ii}$ should be
set to $-3$ for the sites {\em in} the cluster.

Using these methods, the probability of finding different WAC's
has been found analytically \cite{MajumDhar,MaRu}. To do this, one
should compute both $\det\Delta$ and $\det\Delta'$. Then the
probability of finding the cluster $S$ would be
\begin{equation}\label{Pdef}
P(S)=\frac{\det\Delta^{'}_S}{\det\Delta}= \det(1+G\,B_S),
\end{equation}
where $G=\Delta^{-1}$, and $\Delta^{'}_S$ and $B_S$ are the
matrices associated with the cluster $S$. As an example, the
probability of finding a site with height one ($S_0$) is found to
be $P(1)=2(1-2/\pi)/\pi^2\simeq 0.074$. These analytic results
have been confirmed by different simulations
\cite{Erz,Manna,ManGras} though it seems there are some small
disagreements \cite{Mahmoud}. Additionally, by takeing two
clusters of this kind, placing far away from each other, one is
able to compute correlations between these clusters. Taking two
clusters to be $S_0$, separating with the distance $r$, one finds
that the dominant term in correlation functions at large
distances, is proportional to $r^{-4}$ and is independent of
angular distance of the two sites. For other clusters, the radial
dependance is again like $r^{-4}$, but you'll have angular
dependence.

On the other hand, one is interested in the continuum limit of the
problem, that is, finding a field theory that describes the same
model. As ASM is shown to be equivalent to spanning trees and
hence to $q\rightarrow 0 $ of $q$-state Potts  model, the most
appropriate filed theory seems to be $c=-2$ conformal field
theory. Different features of the $c=-2$ model is described in
many articles \cite{c-2}. It has the simple Gaussian action
\begin{equation}\label{FreeAction}
S=\frac{1}{\pi}\int \partial \theta \bar{\partial}\bar{\theta},
\end{equation}
where $\theta$ and $\bar{\theta}$ are complex grassman variables.
Comparing correlation functions, Mahieu and Ruelle \cite{MaRu}
have shown that one can assign a suitable operator in $c=-2$
theory to each WAC in sandpile model. A table of operators versus
clusters for some simple WAC's can be found in \cite{MaRu}.
Actually, they have extended ASM to a model with dissipation in
the following way: the threshold beyond which the column of sand
becomes unstable is increased to $x\,(x>4)$ and during toppling
process $x$ sands are removed from the unstable site, but only
four of them is added to the neighbors. So, $x-4$ grains of sand
are dissipated during each toppling. As a consequence, the action
of $c=-2$ theory should be modified to
\begin{equation}\label{MassiveAction}
S=\frac{1}{\pi}\int\left( \partial \theta
\bar{\partial}\bar{\theta}+ \, \frac{m^2}{4}
\theta\bar{\theta}\right),
\end{equation}
where $m^2$ is equal to $x-4$.

The operators assigned to different WAC's are very similar, in
fact all of them are of the form
\begin{equation}
\phi_S(z)= -\left\{A:\!\d\t \db\tb + \db\t\d\tb \!: + B_1 :\!\d\t
\d\tb + \db\t\db\tb \!: + i B_2 :\!\d\t \d\tb - \db\t\db\tb \!: +
\, C\, P(S) {m^2 \over 2\pi}  :\!\t\tb \!: \right\}. \label{phi}
\end{equation}
with $A$, $B_1$, $B_2$ and $C$ to be determined by comparing with
the results of Dhar and Majumdar method. Also $P(S)$ is the
probability of finding the cluster $S$. The parameter $C$ is
present only in the massive theory and due to Mahieu and Ruelle it
is very striking that $C$ takes only integer values and is equal
to number of sites of the cluster $S$.

In the next section, we will show a more direct way to find these
operators. Also the connection between ASM and $c=-2$ model will
be established more precisely.

\section{Connection Between ASM and $c=-2$ Theory}
The first step to connect the two theories could be considering
their action and partition functions. The $c=-2$ action is given
by equation (\ref{FreeAction}), but in the case of ASM we have not
yet introduce the action. This can be done by considering that in
the steady state there are $\det\Delta$ different configurations
with the same probability. So, the partition function of the model
is just equal to $\det\Delta$. One can write the determinant of an
arbitrary matrix in terms of Gaussian integration on grassman
variables, namely:
\begin{equation}
\det\Delta=\int \prod d\t_i \prod d\tb_j \exp\left(\sum \t_i
\Delta_{ij} \tb_j\right).
\end{equation}
As the matrix $\Delta$ is the discrete version of Laplacian
operator, it is clear that the above action leads to the action
(\ref{FreeAction}) in continuum limit. This was noted before by
Ivashkevich \cite{IvashAction}.

Now we turn on to scaling fields assigned to WAC's. The starting
point is equation (\ref{Pdef}), where the probability of finding
the cluster $S$ is given. Again, the determinant of $\Delta^{'}$
can be written in terms of Gaussian integrals, so the probability
$P(S)$ turns out to be
\begin{equation}
P(S)=\frac{\det\Delta^{'}_S}{\det\Delta}=\frac{\int d\t_i d\tb_j
\exp\left(\sum \t_i \Delta^{'}_{ij} \tb_j\right)}{\int d\t_i
d\tb_j \exp\left(\sum \t_i \Delta_{ij} \tb_j\right)}=\frac{\int
d\t_i d\tb_j \exp\left(\sum (\t_i \Delta_{ij} \tb_j+\t_i B_{ij}
\tb_j)\right)}{\int d\t_i d\tb_j \exp\left(\sum \t_i \Delta_{ij}
\tb_j\right)}.
\end{equation}
Looking carefully at the above relation, one is able to derive the
proper discrete field for the cluster $S$:
\begin{equation}\label{SCField1}
\varphi_S=\exp\left(\sum \t_i B_{ij} \tb_j\right).
\end{equation}
The next step is to translate this field to continuum language.
This translation should be done very carefully. Using the
definition of the matrix $B$, the continuum version of the term
inside the exponential is calculated easily. For the clusters
shown in figure 1, we have
\begin{eqnarray}
\sum \t_i B_{ij} \tb_j |_{S=S_0}  &\propto& \d \t \db \tb+
\db\t\d\tb ,\nonumber\\
\sum \t_i B_{ij} \tb_j|_{S=S_1} &\propto& 3 (\d \t \db \tb+
\db\t\d\tb) - (\d\t\d\tb+\db\t\db\tb)+ \ldots.
\end{eqnarray}
The exponential can be expanded in powers of $\t$ and $\tb$, but
since these variables are grassmans, the series ends at quadratic
term. So, the operators we are interested in, are just the above
fields, apart from an unimportant unity term in the series.

The operator derived here for $S_0$ is in complete agreement with
the one derived in \cite{MaRu}, you should just put the
probability of finding height one to complete the proportionality.
The field obtained for the cluster $S_1$, however, is a little bit
different, but not quite far away. The corresponding operator in
\cite{MaRu} is
\begin{equation}
\varphi_{S_1} \simeq 0.020143 (\d \t \db \tb+ \db\t\d\tb) -
0.006190 (\d\t\d\tb+\db\t\db\tb),
\end{equation}
and the ratio of the two coefficients is about 3.25. Finding other
operators corresponding to other clusters, one observes the same
deviations, that is, although the ratios between different
coefficient are not just the same as the one derived by
\cite{MaRu}, but they are very close to them. This shows that we
are on the right path.

To derive better result, one should do the process of continuation
with more care. This means that we should first expand the
exponential in terms of $\t$ and $\tb$ before going from discrete
to continuum. Let's consider a two point correlation function.
Using Dhar and Majumdar method, the height correlation of two
WAC's have the following form
\begin{eqnarray}
P(S^{o},S^{p})=\det(1+G\,B)&=&\frac{\int d\t_i d\tb_j
\exp\left(\sum (\t_i \Delta_{ij} \tb_j+\t_i B_{ij}^{o} \tb_j+\t_i
B_{ij}^{p} \tb_j)\right)}{\int d\t_i d\tb_j \exp\left(\sum \t_i
\Delta_{ij}
\tb_j\right)}\nonumber\\
\nonumber\\ &=&\left\langle\exp\left(\t_i B_{ij}^{o}
\tb_j\right)\exp\left(\t_i B_{ij}^{p} \tb_j\right)\right\rangle.
\end{eqnarray}
Expansion of the exponentials leads to several terms whose
expectation values can be obtained using Wick theorem. We
categorize the contractions in the following way: firstly, one may
contract all the grassman variables in each of the clusters with
themselves; secondly, only two connections are established between
the two clusters\footnote{The number of connections should be
even, because if you take it to be odd, odd number of $\t$ and
$\tb$ remain in each cluster and so the contribution of such term
is zero} and so on. With every connection between the two WAC's,
we will have a long range Green function. This means that the
terms with more long range connections fall off more rapidly. As
we are interested in the leading terms of the correlations
function, only the first two terms could be considered
\begin{eqnarray}
P(S^{o},S^{p})&=&\left\langle(1+\t_i B_{ij}^{o} \tb_j+...)(1+\t_k
B_{kl}^{p} \tb_l+...)\right\rangle
\nonumber\\&=&P(S^{o})P(S^{p})+\left\langle\left\langle(1+\t_i
B_{ij}^{o} \tb_j+...)(1+\t_k B_{kl}^{p}
\tb_l+...)\right\rangle\right\rangle,
\end{eqnarray}
where $\left\langle\left\langle\cdots\right\rangle\right\rangle$
indicates that only the contractions with two long connections
should be considered. The first term, which comes from
inter-contractions only, is simply equal to product of
probabilities of the two clusters. However, in the second term one
finds nontrivial correlations, and hence could be considered to
contain the scaling operators we are looking for.

The process of derivation of the operator associated with a given
cluster can be summarized as follows. First we should expand the
exponential $\exp(\t_i B_{ij} \tb_j)$. Then contract all the
variables in any possible way, leaving one $\t$ and one $\tb$
uncontracted. As an example if we have a term in the exponential
of the form $\t\tb\t\tb$ we should transform it to
\begin{equation}
\t_1\tb_2\t_3\tb_4\rightarrow \t_1\tb_2G_{34}-\t_1\tb_4G_{32}+
\t_3\tb_4G_{12}-\t_3\tb_2G_{14}.
\end{equation}
Here, $G_{ij}$ is the Green function between the two sites $i$ and
$j$. The last step is to state the obtained expression in terms of
$\t$, $\tb$ at the origin and their derivatives.

Doing all these together one is able to derive the desired scaling
field. Though the procedure is straitforward, it is very long and
cumbersome. But fortunately, the method of expansion has a nice
graphic representation as follows. First we put a point for every
site involved in the matrix $B$. The contraction between
$\theta_i$ and $\tb_j$ is represented with an arrow pointing from
$i$ to $j$, if $i=j$ then we draw a loop. For the two sites whose
variables are not contracted, one draws a dashed line from one to
the other. Then for every line pointing from $i$ to $j$, one puts
the corresponding Green function, $G_{ij}$ multiplied by $B_{ij}$.
For the dashed line, one simply puts $B_{ij}$. The last rule is to
put a minus sign for every loop with even number of lines. For
example for the cluster $S_0$ two of the possible graphs and the
corresponding expressions are shown in figure 2.
\begin{figure}
\begin{picture}(200,140)(0,20)
\includegraphics{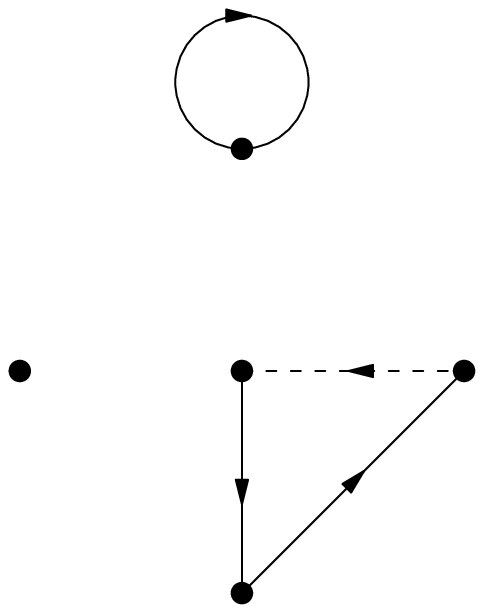} \includegraphics{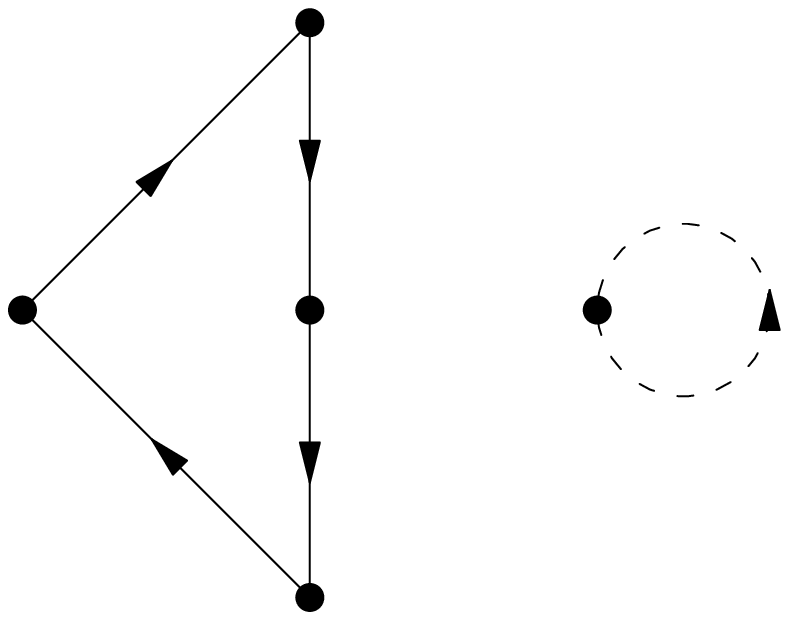} \put(43,95){$L$} \put(90,100){$O$}
\put(140,95){$R$} \put(90,138){$U$} \put(90,44){$D$}
\put(238,95){$L$} \put(281,96){$O$} \put(340,95){$R$}
\put(290,145){$U$} \put(290,44){$D$}
\put(0,23){{\footnotesize$\left(G_{OD}G_{DR}G_{LL}B_{OD}B_{DR}B_{LL}B_{RO}\right)\times\t_R\tb_O$}}
\put(200,23){{\footnotesize$-\left(G_{OD}G_{DL}G_{LU}G_{UO}B_{OD}B_{DL}B_{LU}B_{UO}B_{RR}\right)\times\t_R\tb_R$}}
\end{picture}
\caption{ Two examples of many possible graphs related to the
cluster $S_0$, and their corresponding expressions}
\end{figure}

This equivalency makes the job more tractable. We draw all of the
possible graphs and write the equivalent terms in the series.
Then, By expanding $\tb_i$ and $\theta_j$ in terms of $\t$, $\tb$
and their derivatives, we arrive at the scaling field of the
specific cluster. The result for the cluster $S_0$ is
\begin{equation}
\phi_{S_0}(z)= -\frac{2(\pi-2)}{\pi^{2}}:\!\d\t \db\tb +
\db\t\d\tb : ,
\end{equation}
which, apart from a factor of $\pi$, is the same as the one
derived in \cite{MaRu}. The $1/\pi$ factor, comes in the very same
way as it appears in the action (\ref{FreeAction}).

The result can be confirmed by doing it in another way. If we
contract the remaining $\tb_i$ and $\theta_j$, in the series (to
produce an additional $G_{ij}$), then the series will be equal to
$P(S)$, or equivalently $\det(1+BG)$. So, the term proportional to
$G_{ij}$ in the expression for $\det(1+BG)$ is simply the
coefficient of $\t_i\tb_j$ in the above series. This method is
much faster, specially if done numerically. For other clusters
like $S_1$, the corresponding scaling field is derived by the
latter method and the results are in agreement with \cite{MaRu}.

There remain some points to be clarified. First, we should note
that any correlation function in $c=-2$ theory is zero unless you
have the zero mode $\t\tb$ in the correlation. So all the
correlations like $\langle \phi_{S^o} \phi_{S^p}\rangle$ should be
computed as $\langle \phi_{S^o} \phi_{S^p}\t\tb\rangle$. This has
a physical interpretation. The theory could not be defined on the
whole plane, as the boundaries play an important role in ASM. In
fact we should put mass term on the boundaries so that the grains
of sand have the opportunity to leave the system. But if we would
like to have the whole plane, it would be enough to put a field
$\t\tb$ at infinity.

The other point is about the factor $C$ in equation (\ref{phi}).
By the method we derive the scaling field, it is quite natural
that this factor should be an integer and the integer is just the
number of sites in the cluster. Because in the massive theory, the
matrix $B$ is a little bit changed: $B_{ii}$ is set to $1-x$
instead of -3. So, for each site of the cluster we will have an
additional term proportional to $m^2\, \t\tb$, which is multiplied
by the probability of the cluster when the contraction of other
variables is done.

\end{document}